\documentclass{iaus}
\usepackage{graphicx}

\title[Synthesis models in the VO] 
{Synthesis models in the VO framework}

\author[Cervi\~no et al.]
{M. Cervi\~no$^{1,4}$,
E. Terlevich$^{2}$,  
R. Terlevich$^{2}$,
C. Rodrigo-Blanco$^{3,4}$,
V. Luridiana$^{1,4}$,
A. L\'opez$^{2}$  \and E. Solano$^{3,4}$}
\affiliation{$^1$Instituto de Astrof\'\i sica de Andaluc\'\i a (IAA-CSIC), Camino bajo de Hu\'etor 50, 18008 Spain\break email: mcs@iaa.es, vale@iaa.es\\[\affilskip]
$^3$Instituto Nacional de Astrof\'\i sica \'Optica y Electr\'onica, Puebla, M\' exico\break email: eterlevi@inaoep.mx, rjt@inaoep.mx, allopez@inaoep.mx\\[\affilskip]
$^3$Laboratorio de Astrof\'\i sica Espacial y F\'\i sica Fundamental (LAEFF-INTA), Apdo 50727, Madrid 28080\break email: crb@laeff.inta.es, esm@laeff.inta.es\\[\affilskip]
$^4$Spanish Virtual Observatory (SVO),  {\tt http://svo.laeff.inta.es}
}

\pubyear{2007}
\volume{241}  
\pagerange{119--126}
\date{?? and in revised form ??}
\jname{Stellar Populations as Building Blocks of Galaxies}
\editors{A. Vazdekis \& R. Peletier, eds.}
\begin{document}

\maketitle

\begin{abstract}

The theory interest group in the International Virtual Observatory Alliance (IVOA) has the goal of ensuring that theoretical data and services are taken into account in the IVOA standards process. In this poster we present some of the efforts carried out by this group to include evolutionary synthesis models in the VO framework. In particular we present the VO tool PGos3, developed by the INAOE (Mexico) and the Spanish Virtual Observatory which includes most of public SSP models in the VO framework (e.g. VOSpec). We also describe the problems related with the inclusion of synthesis models in the VO framework and we try to encourage people to define the way in which synthesis models should be described. This issue has implications not only for the inclusion of synthesis models in the the VO framework but also for a proper usage of synthesis models.



\end{abstract}

\firstsection 

\section{The VO and IVOA}

The Virtual Observatory (VO) is a system by which astronomers can produce more robust scientific results by allowing the use of multiple data centers (observational and theoretical ones), and by the development of analysis and visualization tools. The International Virtual Observatory Alliance (IVOA)\footnote{http://www.ivoa.net} has the mission to facilitate the international coordination and collaboration necessary for the development and deployment of the VO.

One of the IVOA interest groups is the {\it Theory interest group} whose goal is to ensure that theoretical data and services are taken into account in the IVOA standard process.

\section{SSP models in the VO}

\begin{figure}
 \begin{center}
 \includegraphics[height=3.3in]{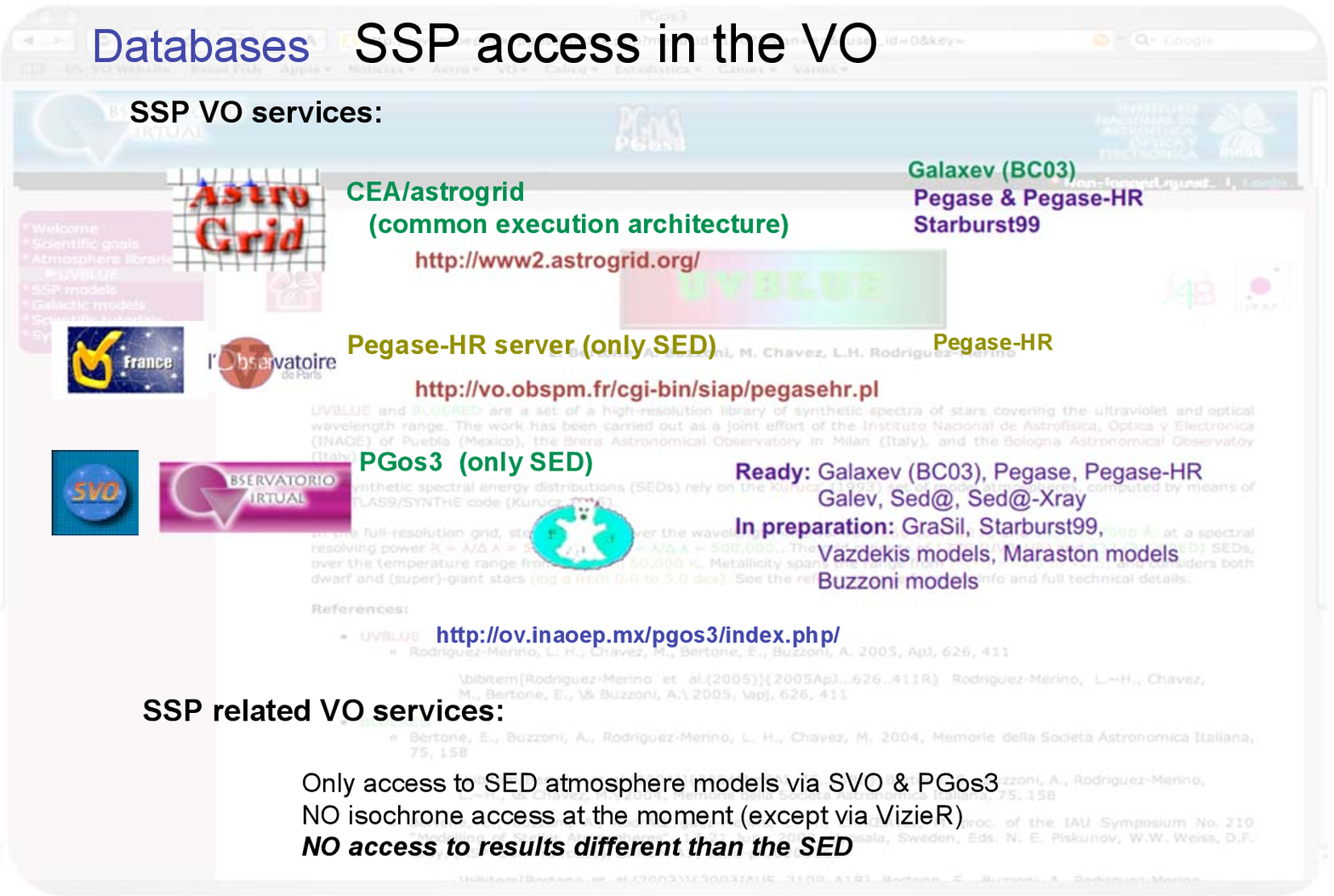}
 \end{center}
  \caption{Schema of current SSP access services in the VO.}
  \label{fig1}
\end{figure}

Currently, it is possible to access synthesis models in the VO framework. However, they are only implemented to access theoretical SEDs via {\it Theoretical Spectral Access Protocol} (TSAP) developed by ESA-VO and SVO, but TSAP does not cover all possible synthesis models results. Moreover, the access is only currently performed by spectral oriented tool (VOSpec\footnote{{\tt http://esavo.esa.int/vospec/}} in particular). Current services are illustrated in Fig. \ref{fig1}. 

\section{The PGos3 database and Synthesis Models description}

PGos3\footnote{{\tt http://ov.inaoep.mx/pgos3/}} is a database aimed to provide SSP and related models in a VO framework for its later use in VO applications. It is included in a long term project for fitting and analysis of stellar populations in galaxies. PGos3 has been developed by the INAOE (Mexico) and the SVO, and it had been used for TSAP development. 

However, to go further in the inclusion of synthesis models in the VO it is necessary to provide a {\it Data Model} of the possible results. That is, it is mandatory: (a) To {\it define} what means a synthesis code result: deterministic vs. probabilistic paradigm like Monte Carlo simulations or general probabilistic descriptions (\cite{CL06}). (b) To define how a synthesis model can be {\it described} univocally in terms of physical/computational parameters and not only code-authors names. And (c) to describe the {\it limitations} and {\it parameter space coverage} of different codes in an homogeneous way.

These requirements are not only VO requirements, but also scientific requirements needed (a) to understand why synthesis models computed by different codes with identical inputs differ, (b) to be able to establish physical justifications when a particular code, instead of a fashionable one, is chosen, and (c) to make a proper use of synthesis models.

\begin{acknowledgments}
This work was supported by the Spanish projects PNAYA2005-24102-E and PNAYA2004-02703. MC is supported by a Ram\'on y Cajal fellowship. VL is supported by a CSIC-I3P fellowship. 
\end{acknowledgments}

\end{document}